\pdfoutput=1
\RequirePackage{ifpdf}
\ifpdf 
\documentclass[pdftex]{sigma}
\else
\documentclass{sigma}
\fi

\usepackage{subfigure}

\numberwithin{equation}{section}

\begin{document}

\allowdisplaybreaks

\renewcommand{\thefootnote}{$\star$}

\renewcommand{\PaperNumber}{110}

\FirstPageHeading

\ShortArticleName{Statistical Thermodynamics of Polymer Quantum Systems}

\ArticleName{Statistical Thermodynamics\\ of Polymer Quantum Systems\footnote{This
paper is a contribution to the Special Issue ``Loop Quantum Gravity and Cosmology''. The full collection is available at \href{http://www.emis.de/journals/SIGMA/LQGC.html}{http://www.emis.de/journals/SIGMA/LQGC.html}}}

\Author{Guillermo CHAC\'{O}N-ACOSTA~$^\dag$, Elisa MANRIQUE~$^\ddag$, Leonardo DAGDUG~$^\S$\\ and Hugo A.~MORALES-T\'ECOTL~$^\S$}

\AuthorNameForHeading{G.~Chac\'{o}n-Acosta, E.~Manrique,  L.~Dagdug and H.A.~Morales-T\'ecotl}

\Address{$^\dag$~Departamento de Matem\'aticas Aplicadas y Sistemas,
Universidad Aut\'onoma\\
\hphantom{$^\dag$}~Metropolitana-Cuajimalpa, Artificios 40,
M\'exico D. F. 01120, M\'exico}
\EmailD{\href{mailto:gchacon@correo.cua.uam.mx}{gchacon@correo.cua.uam.mx}}

\Address{$^\ddag$~Institut f\"{u}r Physik,
Johannes-Gutenberg-Universit\"{a}t, D-55099 Mainz, Germany}
\EmailD{\href{mailto:manrique@zino.physik.uni-mainz.de}{manrique@zino.physik.uni-mainz.de}}

\Address{$^{\S}$~Departamento de F\'\i sica, Universidad
Aut\'onoma Metropolitana-Iztapalapa,\\
\hphantom{$^\S$}~San Rafael Atlixco 186,
M\'exico D. F. 09340, M\'exico}
\EmailD{\href{mailto:dll@xanum.uam.mx}{dll@xanum.uam.mx}, \href{mailto:hugo@xanum.uam.mx}{hugo@xanum.uam.mx}}

\ArticleDates{Received September 01, 2011, in f\/inal form November 16, 2011;  Published online December 02, 2011}

\Abstract{Polymer quantum systems are mechanical models quantized
similarly as loop quantum gravity. It is actually in quantizing
gravity that the polymer term holds proper as the quantum geometry
excitations yield a reminiscent of a polymer material. In such an approach both
non-singular cosmological models and a microscopic basis for the
entropy of some black holes have arisen. Also important physical questions for these systems
involve thermodynamics. With this motivation, in this work, we study the statistical
thermodynamics of two one dimensional {\em polymer} quantum systems:
an ensemble of oscillators that describe a solid and
a bunch of non-interacting particles in a box, which thus form
an ideal gas.
We f\/irst study the spectra of these polymer systems. It turns
out useful for the analysis to consider the length scale required by
the quantization and which we shall refer to as polymer length. The
dynamics of the polymer oscillator can be given the form of that for
the standard quantum pendulum. Depending on the dominance of the polymer
length we can distinguish two regimes: vibrational and rotational.
The f\/irst occur for small polymer length and here the standard
oscillator in Schr\"{o}dinger quantization is recovered at leading
order. The second one, for large polymer length, features dominant
polymer ef\/fects. In the case of the polymer particles in the box,
a bounded and oscillating spectrum that presents a band structure and a Brillouin
zone is found.
The thermodynamical quantities calculated with these spectra have
corrections with respect to standard ones and they depend on the
polymer length. When the polymer length is small such corrections
resemble those coming from the phenomenological generalized
uncertainty relation approach based on the idea of the existence of
a minimal length. For generic polymer length, thermodynamics of both
systems present an anomalous peak in their heat capacity~$C_V$. In
the case of the polymer oscillators this peak separates the
vibrational and rotational regimes, while in the ideal polymer gas
it ref\/lects the band structure which allows the existence of
negative temperatures.}


\Classification{82B30; 81S05; 81Q65; 82B20; 83C45}

\renewcommand{\thefootnote}{\arabic{footnote}}
\setcounter{footnote}{0}

\section{Introduction}
\vspace{-2mm}

In coping with the challenge of quantizing gravity the loop
quantization approach~\cite{rov,thiemann} has proved convenient to
incorporate the background independent character demanded by general
relativity. Important progress in this approach include the
avoidance of the classical singularity which in loop quantum
cosmology is replaced by a quantum bounce~\cite{bojoLR} in physically motivated models,
\cite{bojoPRL,bojoGRG,apslqc}, and a microscopic
basis for the entropy of some black holes that is in accordance with
the Bekenstein--Hawking's semiclassical formula~\cite{rovbh,abckBH,doma}.

\looseness=-1
However some further physical questions in regard to cosmology and
black holes necessarily involve thermodynamics. As it is well known
primordial particle backgrounds including neutrinos, gravitons and
photons, originating in the early universe provide windows to
explore such early stage \cite{kolb,weing}. For instance the
stochastic graviton remnant and its statistical properties have been
studied \cite{GS1,GS2}. Also the dif\/ferent contributions to the
spectrum of gravitons produced during the super-inf\/lationary period
have been investigated \cite{Mie1,Mie2,bojohoss,Mie3}. In regard to black holes, from the loop quantum gravity
perspective, the challenge remains of describing black hole
evaporation including in particular its thermodynamical aspects (see e.g.~\cite{BHevap-Ashtekar} for recent work.)

Rather than dealing with the thermodynamics of loop quantized
gravitational systems a more tractable problem is to consider the statistical thermodynamics
of polymer quantum systems \cite{ashpqm,CVZ1,CVZ}. The latter are
mechanical systems quantized following loop quantum
gravity. Here an important comment in regard to the polymer term is
in order. In the gravitational case, loops, or more strictly, graphs,
label quantum states of gravity. This yields a picture resembling
polymer materials, which justif\/ies adopting the term polymer
quantization for gravity. However, as we will see below, for
mechanical systems states will be labeled by point sets belonging to
a~lattice. Thus, although the term polymer is inherited from the
gravity case it is not actually  realized in the mechanical one. We
should stress at this point also that in loop or polymer
quantization a  length scale is required for its construction, while
for the gravitational case this is identif\/ied with Planck's length,
in the mechanical case it is just a free parameter and we refer to
it as the polymer length scale. It should be mentioned though that,
technically, the quantization here dubbed polymer, was previously
considered from a dif\/ferent perspective in the form of a non regular representation of the
canonical commutation relation~\cite{Prev}. Interestingly, a quantization based on dif\/ference operators has also been considered~\cite{Dobrev}.

Polymer quantum systems have been convenient to illustrate some
features arising in loop quantum gravity~\cite{ashpqm,fred}. In
particular they have the same conf\/iguration space as that of loop
quantum cosmology~\cite{vel}. The continuum limit of the polymer
quantum system has also been explored using ad hoc renormalization
schemes in which the polymer length scale runs~\cite{CVZ1,CVZ}. Investigating whether polymer quantum system
admits Galilean symmetry has been reported in~\cite{chiou}.
Inspired by the cosmic singularity avoidance this quantization has
been used to explore potentials such as~$1/r$~\cite{Huss}, and~$1/r^2$~\cite{Kunst}.

In this work we shall study the thermostatistics of two simple
polymer systems, namely, an ensemble of oscillators and a bunch of
noninteracting particles in a box. The paper is structured as
follows. In Section~\ref{section2} we review the basics of polymer
quantization of mechanical systems, in particular the eigenvalue
problem. For the harmonic oscillator the corresponding eigenvalue
problem can be casted in Fourier space as a second order
dif\/ferential equation. In this way one can see its spectrum is just that of
the standard quantum pendulum. Two regimes can be seen to appear: an
oscillatory or vibrational one and a rotational one. In the f\/irst, the oscillator in Schr\"{o}dinger quantization is recovered,
at leading order, while in the second regime, the polymer ef\/fects
are dominant. Furthermore, we solve exactly the eigenvalue problem
for the polymer particle in a box that takes the form of a second
order dif\/ference equation. We obtain a bounded and oscillating
spectrum that features a band structure and a Brillouin zone.

In Section~\ref{section3} we calculate the corresponding thermodynamical
quantities with these spectra. They depend on the
polymer length and behave dif\/ferently from those in standard
thermodynamics, which is recovered when the polymer length is
considered small. In this case the thermodynamical variables can be written as
a series on the small polymer length, similarly to what happens in the
phenomenological Generalized Uncertainty Principle (GUP) approach
based on the idea of the existence of a minimal length
\cite{gacnat,gac}. These were applied previously to an ideal
gas \cite{kal,nozmeh,noz}, and radiation
\cite{nozsef,camacho}. (The modif\/ication to the statistical
mechanics of systems were also studied from the perspective of the
extension to the Standard Model that have Lorentz violating terms
\cite{don}, and the case of radiation was also studied with
corrections arising from loop quantum gravity~\cite{ipny,amtu2}.) In some sense this behavior could be expected since the
uncertainty relation for polymer systems is quite similar to GUP
\cite{ashpqm,PUR}. We shall show quadratic corrections occur in the energy of the polymer particle in a box, just as in GUP, but with opposite sign.
For a generic, not necessarily small, polymer length, thermodynamics of both
systems present an anomalous peak in their heat
capacity $C_V$. For the polymer oscillators this peak separates the
vibrational and rotational regimes, while in the ideal polymer gas
it ref\/lects the band structure. It is worth stressing that, for
the gas, the band structure allows for the existence of negative
temperatures.

Finally in Section~\ref{section4} we discuss our results and point out some
perspectives.

\section{Eigenvalue problem in the polymer representation\\ of quantum mechanics}\label{section2}

In this section we describe the main features of the polymer
quantization of a non relativistic particle moving on the real line
\cite{ashpqm,CVZ1}, and describe brief\/ly the corresponding
eigenvalue problem. To do so we start by noticing that instead of
the Heisenberg algebra, involving posi\-tion~$\hat{q}$ and momentum
$\hat{p}$ of the particle
\begin{gather*}
    [\hat{q},\hat{q}]=[\hat{p},\hat{p}]=0,
\qquad
    [\hat{q},\hat{p}]=\hat{q}\hat{p}-\hat{p}\hat{q}=i \hat{I},
\end{gather*}
where $\hat{I}$ is the identity on Hilbert space $\mathcal{H}$, one
adopts the Weyl algebra
\begin{gather*}
  U(\lambda_1) \cdot U(\lambda_2) = U(\lambda_1 + \lambda_2), \qquad 
  V(\mu_1) \cdot V(\mu_2) = V(\mu_1 + \mu_2), \\ 
  \hat{U}(\lambda) \cdot \hat{V}(\mu) = e^{-i\lambda \mu}
    \hat{V}(\mu) \cdot \hat{U}(\lambda). 
\end{gather*}
According to Stone--von Neumann theorem the elements of the above
algebras can be related under certain conditions, including in
particular weak continuity. Namely
\begin{gather*}
    \hat{U}(\lambda) = e^{i\lambda \hat{q}}, \qquad \hat{V}(\mu) = e^{i\mu \hat{p}},
\end{gather*}
which however, does not hold in the polymer case.

In the usual or \emph{Schr\"{o}dinger} representation of quantum
mechanics the Hilbert space is $\mathcal{H}=L^2(\mathbb{R},dx)$ with
the Lebesgue measure $dx$. Instead, in the \textit{loop or polymer}
representation the kinematical Hilbert space $\mathcal{H}_{\rm poly}$ is
the Cauchy completion of the set of linear combination of some basis
states $\{|x_j\rangle\}$, whose coef\/f\/icients have a suitable
fall-of\/f \cite{ashpqm}, and with the following inner
product
\begin{gather*}
    \langle x_i|x_j\rangle = \lim_{T\rightarrow \infty}
\frac{1}{T}\int_a^{a+T}dk\, e^{ik(x_i-x_j)} =\delta_{x_i, x_j},
\end{gather*}
where $\delta_{x_i, x_j}$ is the Kronecker delta, instead of Dirac
delta as in Schr\"{o}dinger representation, then we say that the
orthonormal basis is discrete. The kinematical Hilbert space can be
written as
$\mathcal{H}_{\textrm{\rm poly}}=L^2(\mathbb{R}_d,\textrm{d}\mu_{d})$
with $d\mu_d$ the corresponding Haar measure, and $\mathbb{R}_d$ the
real line endowed with the discrete topology\footnote{In the
\textit{momentum} representation, the conf\/iguration space is the
Bohr compactif\/ication of the real line $\mathbb{R}_B$, \cite{ashpqm,vel}.}.

As we already notice Stone--von Neumann's theorem is no longer
applicable since the opera\-tor~$V(\mu)$ fails to be weakly continuous
on the parameter $\mu$ due to the discrete structure assigned to
space \cite{funal,ashpqm}. The shift operator $\hat{V}(\mu)$
is not related to any Hermitian operator as inf\/initesimal generator.
Hence, for the representation of the Weyl algebra we choose the
position operator~$\hat{x}$ and the translation~$\hat{V}(\mu)$
instead of the momentum operator. Its action on the basis~is:
\[
\hat{x}|x_j\rangle=x_j|x_j\rangle,\qquad
\hat{V}(\mu)|x_j\rangle=|x_j-\mu\rangle,
\]
fulf\/illing also:
\begin{gather*}
    [\hat{x},\hat{V}(\mu)]=-\mu\hat{V}(\mu).
\end{gather*}

Since there is no well def\/ined momentum operator, any function on
phase space which depends on the momentum has to be regularized. In
particular this is so for the Hamiltonian. To do so we introduce an
extra structure, namely a regular lattice with spacing length~$\mu_0$. This analogue of what happens in Loop Quantum Cosmology,
where there is a fundamental minimum area
\cite{bojoLR,bojoPRL,bojoGRG,apslqc} given in terms of Planck length.

Let us consider $\mu_0>0$ as any f\/ixed scale. In general $\mu_0$ can be function of $x$ but here
we suppose it is constant. One of the simplest options to def\/ine an
operator analogous to the momentum operator is:
\begin{gather}\label{K}
    \hat{K}_{\mu_0} = \frac{1}{2i\mu_0}\bigl(\hat{V}(\mu_0)-\hat{V}(-\mu_0)\bigr).
\end{gather}
This choice can be thought of as the formal replacement
\[\hat{p} \rightarrow \frac{1}{\mu_0} \widehat{\sin{(\mu_0 p)}},\]
where the right hand side is given by (\ref{K}). Some authors have
studied a semiclassical regime in which the expectation value of the Hamiltonian is taken with respect to a semiclassical state to yield an ef\/fective dynamics that can be seen, at leading order, as obtained from the replacement
$p\rightarrow \sin{(\mu_0 p)}/\mu_0$ in the classical Hamiltonian. This
has been proved useful in several models
\cite{Poly1,Poly2,Poly3,taub,taub0}.

Thus the polymer Hamiltonian is written as
\begin{gather}\label{ham}
    \widehat{H}_{\mu_0} = \frac{\hbar^2}{2m\mu_0^2} \big[ 2- \hat{V}(\mu_0) -\hat{V}(-\mu_0)\big]+ \hat{W}(x),
\end{gather}
where\looseness=-1 \ $W(x)$ is a potential term. The dynamics generated by
(\ref{ham}) decomposes the polymer Hilbert space
$\mathcal{H}_{\textrm{\rm poly}}$, into an inf\/inite superselected
f\/inite-dimensional subspaces, each with support on a regular lattice
$\gamma=\gamma(\mu_0,x_0)$ with the same space between points
$\mu_0$, where $\gamma(\mu_0,x_0) = \{n \mu_0 + x_0\,|\,n\in
\mathbb{Z}\}$, and $x_0\in[0,\mu_0)$. Thus, choosing $x_0$ f\/ixes the
superselected sector.

There are at least two possible ways to regain Schr\"{o}dinger
quantum mechanics. One is to consider the polymer
length scale $\mu_0$ as small such that the dif\/ference equation~(\ref{ham})
can be approximated by the usual dif\/ferential Schr\"{o}dinger
equation~\cite{ashpqm}. Another possibility is to consider the
introduction of the lattice as an intermediate step, after which it
is necessary to carry out a renormalization procedure, as it is usually
done in lattice theories~\cite{creutz}. In polymer quantization the
f\/inal result turns out to be the usual Schr\"{o}dinger quantum
mechanics~\cite{CVZ,CVZ1}.

Next we consider the eigenvalue problem:
\begin{gather}\label{eigen}
    \widehat{H}_{\mu_0}|\psi\rangle=E|\psi\rangle.
\end{gather}
In the case of the Hamiltonian (\ref{ham}) on the lattice
$\gamma(x_0,\mu_0)$, any state $|\psi\rangle\in\mathcal{H}_{\rm poly}$
is of the form:
\begin{gather}\label{3}
    |\psi\rangle=\sum_{j\in\mathbb{Z}}\psi(x_{0}+j\mu_0)|x_{0}+j\mu_0\rangle.
\end{gather}

The substitution of (\ref{3}) into (\ref{eigen}) gives a dif\/ference
equation in the position representation
\begin{gather}\label{3.1}
    \frac{\hbar^2}{2m\mu_0^2} \left[ 2\psi(x_j) - \psi(x_j+\mu_0) - \psi(x_j-\mu_0)\right]= \left[E-W(x_j) \right]\psi(x_j).
\end{gather}
Using the Fourier transform def\/ined by
\begin{gather*}
    \psi(k)=(k|\psi\rangle=\sum_{j\in\mathbb{Z}}\psi(x_{0}+j\mu_0)e^{-ik(x_{0}+j\mu_0)},
\end{gather*}
where $k\in[-\pi/\mu_0,\pi/\mu_0]$ and with $\psi(k)$
satisfying  the condition
\begin{gather*}
    \psi\left(\frac{\pi}{\mu_0}\right)=e^{-2\pi i\frac{x_0}{\mu_0}}\psi\left(-\frac{\pi}{\mu_0}\right),
\end{gather*}
the equation (\ref{3.1}) reads:
\begin{gather}\label{6}
     \Bigg(1-\frac{\mu_0^{2}m}{\hbar^2}E-\cos{k\mu_0}\Bigg)\psi(k) =-\frac{\mu_0^{2}m}{\hbar^2}\sum_{j\in\mathbb{Z}}\psi(x_{0}+j\mu_0)W(x_{0}+j\mu_0)e^{-ik(x_{0}+j\mu_0)}.
\end{gather}
In our analysis two specif\/ic cases will be considered: the harmonic
oscillator and the particle in a box.

\subsection{Harmonic oscillator}\label{section2.1}

The polymer harmonic oscillator has been already studied in
\cite{ashpqm}  and \cite{Hoss-Osc}\footnote{Actually a previous
study appears in \cite{DHO}, in a dif\/ferent context, transport of
electrons in semiconductors.}, using the potential ${W}(x)=\hbar
\omega x^2/2d^2$, with $d^2 = \hbar/m\omega$ a characteristic length
of the oscillator. After performing the Fourier transform in the
r.h.s.\ of (\ref{6}) and using the quadratic potential, one obtains a~second order dif\/ferential equation which can be recognized as a
Mathieu equation \cite{abrwtz}
 \begin{gather}\label{Mathieu}
    \frac{d^2 \psi(\phi)}{d\phi^2} + \left( a - 2q \cos{2\phi} \right)\psi(\phi) =
    0,
 \end{gather}
where $\phi=\frac{k\mu_0+\pi}{2}$, $a = \frac{8}{\lambda^4}\left(
\frac{\lambda^2}{\hbar \omega}E-1 \right)$ and $q = 4\lambda^{-4}$,
and we introduce $\lambda := \mu_0/d$ as a dimensionless length
parameter. This system is just a quantum pendulum in $k$ space~\cite{CVZ1}. For low energies one recovers the harmonic oscillator
behavior, while for high energies it becomes a free rigid rotor
\cite{baker}. Equation~(\ref{Mathieu}) has periodic solutions for
particular values of $a=a_n,b_n$ called the Mathieu characteristic
functions, that depend on~$n$,~\cite{abrwtz}. The corresponding wave
functions can be written in terms of the Mathieu elliptic sine and
cosine~\cite{Hoss-Osc}. The energy eigenvalues can be expressed  as
follows \cite{abrwtz,Hoss-Osc}
\begin{gather}
  E_{2n}  =   \frac{\hbar\omega}{\lambda^2}\left[1 + \frac{\lambda^4}{8}a_n\left(\frac{4}{\lambda^4}\right)\right],\label{osca} \\
  E_{2n+1}  =   \frac{\hbar\omega}{\lambda^2}\left[1 +
  \frac{\lambda^4}{8}b_{n+1}\left(\frac{4}{\lambda^4}\right)\right].\label{oscb}
\end{gather}
An asymptotic expansion for the characteristic functions, assuming
$\mu_0\ll d$ yields that the energy spectrum $E_{2n}\approx
E_{2n+1}$, can be approximated as \cite{ashpqm,Hoss-Osc}
\begin{gather}\label{t1}
    E_n = \left(n+\frac{1}{2}\right)\hbar \omega - \left(\frac{2n^2+2n+1}{32} \right)\lambda^2\hbar \omega +
    \mathcal{O}\big(\lambda^4\big).
\end{gather}
As expected the spectrum consists of the standard harmonic oscillator
plus corrections of order~$\mathcal{O}(\lambda^2)$. However, this
spectrum is not bounded from below. To enforce correspondence with the Schr\"{o}dinger quantization, we can
ask that the leading term be larger than the f\/irst
correction. This gives us a maximum $n_{\max}$ that depends on
$\lambda$,
\begin{gather}\label{rvo}
    n_{\max}\approx \lambda^{-2}.
\end{gather}
This means that, for small $ \lambda$, we can not probe the spectrum with values of $n$
greater than those allowed by (\ref{rvo}). As in \cite{ashpqm} with
the values of $\mu_0 = 10^{-19}m$, corresponding to the maximum
experimental attainable energy today, and with $d=10^{-12}m$ for
the carbon monoxide molecule, one gets $\lambda = 10^{-7}$. In this case $n_{\max}\simeq 10^{14}$.
This is consistent with \cite{CVZ} and \cite{CVZ1} where a~cut-of\/f in
the energy eigenvalues was introduced that depends on the regulator
scale in order to perform the renormalization procedure that is
necessary to implement a continuum limit of the theory.

On the opposite limit for $\lambda\gg 1$ the eigenstates are
$E_{2n}\approx E_{2n-1}$, for $n=1,2,\ldots$
\begin{gather}\label{t01}
    E_n = \frac{\hbar\omega}{\lambda^2} + \hbar\omega\frac{\lambda^2}{8}n^2 + \mathcal{O}\big(\lambda^{-6}\big),
\end{gather}
where the ground state depends only on $\lambda^{-2}$, and actually
falls of\/f as $\lambda$ increases. As pointed out in \cite{Hoss-Osc} this
case may be relevant for the cosmological constant problem. For the
excited states $n \neq 0$, the $\lambda^2$ term dominates. In this
regimen it is better to interpret $\lambda$ as a ratio of energies.
Indeed the dimensionless parameter $\lambda^2 =
\frac{E_{\rm osc}}{E_{\rm poly}}$, where $E_{\rm osc}=\hbar \omega$ and
$E_{\rm poly}=\frac{\hbar^2}{m\mu_0^2}$ that corresponds to the
coef\/f\/icient of the kinetic term in (\ref{3.1}). Then this regime is
the case of $E_{\rm osc}\gg E_{\rm poly}$; if we consider the polymer length
to be the Planck length then this case corresponds to the
trans-planckian region relevant for both inf\/lationary
cosmology and black holes \cite{TP}.

Following the comparison with the quantum rigid rotor
\cite{Doncheski}, we recall that the spectrum of a quantum rigid
rotor in two dimensions is $\frac{n^2\hbar^2}{2I}$, with
$I=\frac{mR^2}{2}$ being the moment of inertia and $R$ the radius of
the rotor. Notice that the second term in (\ref{t01}) has the same
functional dependence on $n$ as for the rotor with ef\/fective moment
of inertia $I_{{\rm ef\/f}} = \frac{(2\hbar/\omega)^2}{m\mu_0^2}$ and
$R_{{\rm ef\/f}}=\frac{2\sqrt{2}}{\lambda}d$.

\subsection{Particle in a box}\label{section2.2}

The next example is a particle conf\/ined in a box of size $L=N\mu_0$.
The free polymer particle was f\/irst studied in \cite{CVZ1}. Here however we conf\/ine it to a box. In this
case, instead of working in Fourier space we can use directly the
dif\/ference equation~(\ref{3.1}). As in the standard case the
potential is def\/ined as
\begin{gather}\label{pwpot}
    W(x_j)=
           \begin{cases}
             0, & \hbox{$x_0<x_j<x_0+L$}, \\
             \infty, & \hbox{otherwise},
           \end{cases}
\end{gather}
where $x_0\in [0,\mu_0)$. This potential is then realized through
appropriate boundary conditions over elements of
$\mathcal{H}_{\rm poly}$. The particle behaves as a free particle inside
the box and vanishes outside, namely
\begin{gather}\label{bcx}
    \psi(x_0)=\psi(L+x_0)=0, \qquad \forall\, x_0 \in [0,\mu_0).
\end{gather}
We can conveniently rewrite equation (\ref{3.1}) by using
$x_j=x_0+j\mu_0$, as
\begin{gather}\label{J40}
    \psi(j+2)-\left(2-\frac{2mE\mu_0^2}{\hbar^2}\right)\psi(j+1)+\psi(j) = 0.
\end{gather}
Following \cite{elaydi}, we propose the solution of the dif\/ference
equation of second order (\ref{J40}), to be
\begin{gather}\label{d2}
    \psi(j) = a_1r_1^j + a_2r_2^j,
\end{gather}
where $a_i$ are constants coef\/f\/icients and $r_i$ are the roots of
the \emph{characteristic equation}:
\begin{gather}\label{d3}
    r^2-\left(2-\frac{2mE\mu_0^2}{\hbar^2}\right)r+1=0.
\end{gather}
The solutions of (\ref{d3}) are
\begin{gather*}
   r_{\pm} =\left(1-\frac{mE\mu_0^2}{\hbar^2}\right) \pm \frac{1}{2}\sqrt{\frac{8mE\mu_0^2}{\hbar^2}\left(\frac{mE\mu_0^2}{2\hbar^2}-1\right) }.
\end{gather*}
For positive energies, the argument of the square root gives us a
relation between energy and ${2\hbar^2}/{m\mu_0^2}$. If $E\geq
{2\hbar^2}/{m\mu_0^2}$, then the roots $r_{\pm}$ are real numbers\footnote{One of these cases give rise to degenerate roots, such
that the solution (\ref{d2}) changes for $\psi(j) = r^j(a_1 +
a_2j).$}, but incompatible with the boundary conditions, therefore
they yield the trivial solution $\psi(x_j)=0$. This leads to the
idea that the minimum length scale $\mu_0$ imposes a cut-of\/f on the
energy, and hence, energies greater than the cut-of\/f are unphysical. Thus, the only meaningful physical case is when $E<
{2\hbar^2}/{m\mu_0^2}$ which gives us complex roots for $r_{\pm}$.
The solution is given by:
\begin{gather}\label{d5}
    \psi(j) =  C_1 \sin(j\theta) + C_2 \cos(j\theta),
\end{gather}
which is a parametrization of (\ref{d2}) in polar coordinates,
\begin{gather*}
  \cos\theta  =  \left(1-\frac{mE\mu_0^2}{\hbar^2}\right), \qquad
  \sin\theta  =
  \frac{1}{2}\sqrt{\frac{8mE\mu_0^2}{\hbar^2}\left(\frac{mE\mu_0^2}{2\hbar^2}-1\right)}.
\end{gather*}
Imposing the boundary conditions (\ref{bcx}) in (\ref{d5}) we f\/ind
that $\theta = n\pi\lambda$. In this case $\lambda = \mu_0/L = 1/N$
and $n \in \mathbb{Z}$. The eigenfunction of (\ref{J40}) turns out
to be
\begin{gather}\label{wf}
    \psi_n(x_j)  = C \sin{\left(n\pi \lambda
    \frac{x_j}{\mu_0}\right)}=C \sin{\left(n\pi
    \frac{j}{N}\right)}, \qquad 0<j<N,
\end{gather}
where $C$ is a normalization factor
\begin{gather*}
    C = \left[ \sum_{j=0}^N \sin^2{\left(n\pi\lambda j\right)} \right]^{-\frac{1}{2}} = \sqrt{\frac{2}{N}}.
\end{gather*}
With (\ref{wf}) we can write the eigenstate (\ref{3}) as
\begin{gather}\label{333}
    |\psi_n\rangle = \sqrt{\frac{2}{N}} \sum_{x_j} \sin{\left(n\pi \lambda \frac{x_j}{\mu_0}\right)}
    |x_{j}\rangle.
\end{gather}
We notice that the sum in (\ref{333}) goes form $j=0, \ldots, N$.
Since $n$ is an integer it is clear that we can not
build the $n+1$ eigenstate because it would depend on the
previous $n$ states. Thus $0<n<N$. The corresponding energy
spectrum is found to be bounded~\cite{gallinar}
\begin{gather}\label{d9}
    E_n = \frac{\hbar^2}{m\mu_0^2}\left( 1-\cos{n\pi\lambda}
    \right), \qquad n\in\{1,2,\ldots,N-1\}.
\end{gather}
The energy spectrum (\ref{d9}) resembles the tight binding model for
particles in a periodic potential that is not inf\/initely high and
allow tunneling with the nearest neighbors sites
\cite{sakurai,CMP,ISSP}:
\begin{gather*}
    E(\kappa) = E_{*}-2\Delta \cos{\kappa\mu_0},
\end{gather*}
where $E_{*}=\frac{\hbar^2}{m\mu_0^2}$ and $\Delta$ represents the
relevant non diagonal terms that give rise to the energy band,
$\kappa$ is interpreted as a wave vector which take values on
$[-\frac{\pi}{\mu_0},\frac{\pi}{\mu_0}]$, and which, by choosing
periodic boundary conditions \cite{CMP,ISSP}, it takes
discrete values $\kappa = \frac{\pi m}{L}$ with $N\in \mathbb{Z}$
and $-N\leq m\leq N$.

\begin{figure}[t]
\centering
  \includegraphics[width=3.8in]{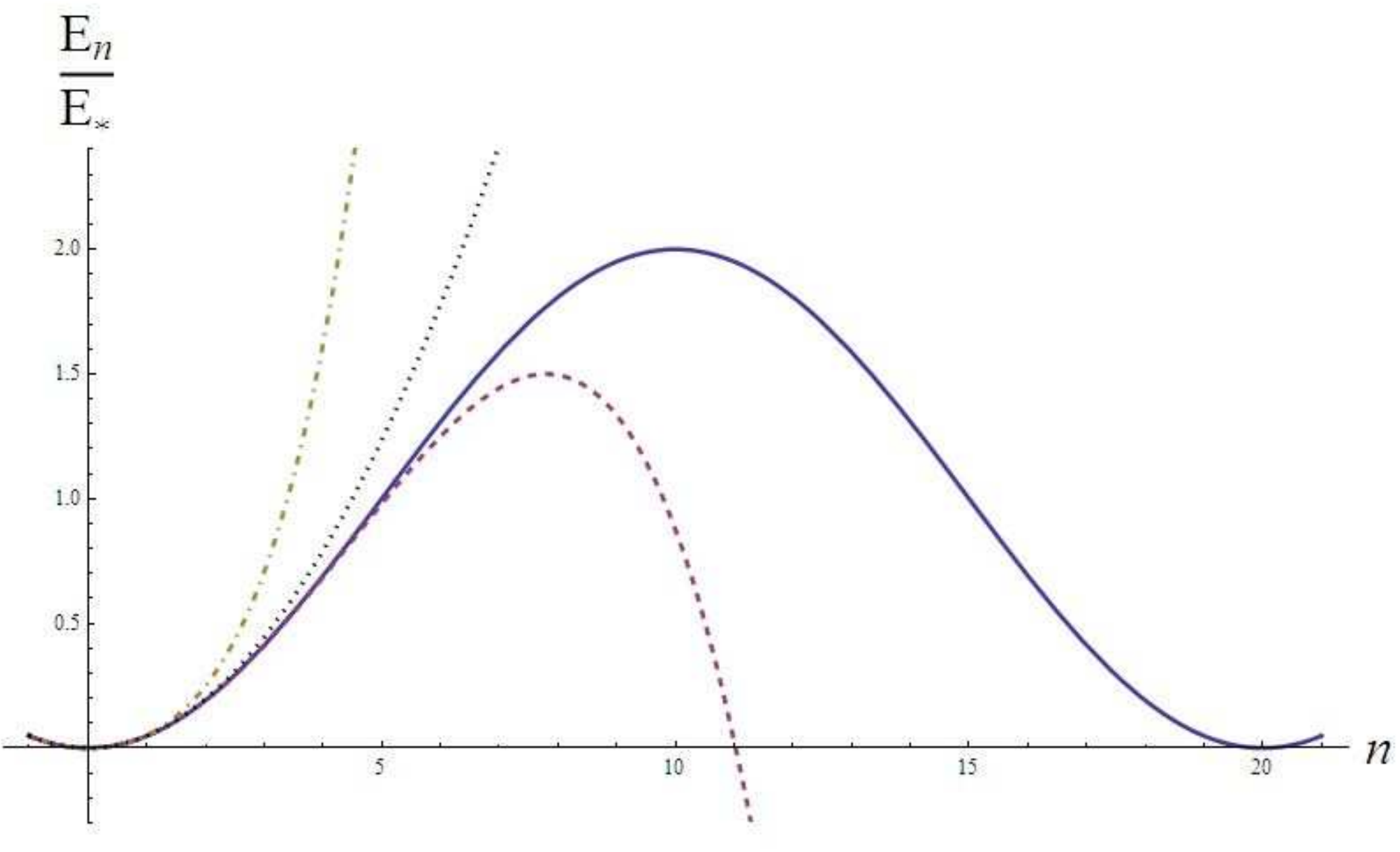}
  \caption{The solid line (blue) corresponds to the f\/irst Brillouin zone for the spectrum
  (\ref{d9}) and the dashed line (red) is the second order approximation equation~(\ref{d10}). The dot-dashed line (green) is the case reported in the GUP literature corresponding to equation~(\ref{d12}) which presents and opposite tendency with respect to the previous one. The dotted line (black) corresponds to the standard case $\lambda = 0$.}\label{BrillSpec}
\end{figure}

Hence, the polymer particle in a box is analogous to the tight
binding model of a particle in a periodic potential with periodic
boundary conditions, having band energy $\Delta=E_{*}/2$. So there
is an energy band appearing in this polymer system given by a energy
spectrum bounded from above and below, and a kind of
\textit{Brillouin zone} as shown in Fig.~\ref{BrillSpec}.

If we expand $\cos{n\pi\lambda}$, for $\lambda \ll 1$, i.e.\ large
size of the box as compared to the lattice spacing, up to second order, we get
 \begin{gather}\label{d10}
    E_n = \frac{\hbar^2n^2\pi^2}{2mL^{2}} - \frac{\hbar^2n^4\pi^4}{24mL^{2}}\lambda^2 + \cdots,
\end{gather}
Using $\mu_0=10^{-19}m$ and with the approximated spectrum (\ref{d10}), the correction will be
signif\/icant only when $n \approx 10^{17}$.

The energy spectrum of a quantum particle in a box has also been
studied in the framework of a GUP \cite{nozazizi}. In this frame the modif\/ications
induced by a minimum length $\ell_{\min}$ on the wave function and
the energy spectrum of a particle in a one-dimensional box, result
in a modif\/ication of the Schr\"{o}dinger equation which transforms
it in a fourth order dif\/ferential equation. The energy spectrum has
a correction proportional to the squared of the minimum length
$\ell_{\min}^2$
\begin{gather}\label{d12}
    E^{\rm (GUP)}_n = \frac{n^2\pi^2\hbar^2}{2mL^2} + \frac{\ell_{\min}^2}{L^2}\frac{n^4\pi^4\hbar^2}{3mL^2}.
\end{gather}
This result can be compared with (\ref{d10}). We can see that the
dependence on the minimum scale is quadratic in both cases but they feature a sign dif\/ference thus signaling opposite tendencies. As we already
mentioned this coincidence is not surprising since polymer systems
have  similar modif\/ications to GUP in their corresponding
uncertainty relation~\cite{PUR}. However, in~\cite{nozazizi} the
usual boundary condition for a particle in a box is used; here,
since the measurement of the position has some uncertainty
proportional to the minimum length, the position
of the walls is not determined precisely. This problem is solved
recalling that in polymer quantum mechanics the dynamics is def\/ined
on superselection spaces for which there is, in each one, a proper
boundary conditions def\/ined by~(\ref{pwpot}) and~(\ref{bcx}).

\section{Polymer corrections to thermodynamic quantities\\ of simple systems}\label{section3}

In standard statistical mechanics \cite{path}, the canonical
partition function $Z$ is def\/ined as the sum over all possible
states
\begin{gather}\label{t2}
    Z(\beta)= \sum_{n}\exp{\left(-\beta E_n\right)},
\end{gather}
where $\beta=(kT)^{-1}$, $k$ is  the Boltzmann's constant, and
$T$ is the temperature. With it we can calculate all thermodynamical
quantities of the system, through the def\/inition of the Helmholtz
free energy \cite{path}
\begin{gather}\label{t7}
    F = -\frac{\mathcal{N}}{\beta} \ln{Z},
\end{gather}
where $\mathcal{N}$ is the particle number. The relationship with
other thermodynamic quantities such as the equation of state,
entropy and chemical potential can be obtained by the standard relations
\begin{gather}
  p = -\frac{\partial F}{\partial L}, \label{t9b} \\
  \mu = \frac{\partial F}{\partial \mathcal{N}}, \label{t9a}\\
  S = k\beta^2\frac{\partial F}{\partial \beta}. \label{t9c}
\end{gather}
Moreover the internal energy and the heat capacity can also be related to the partition function as
relations
\begin{gather}
  U  =  -\mathcal{N} \frac{\partial \ln{Z}}{\partial \beta} = -\frac{\mathcal{N}}{Z} \frac{\partial Z}{\partial \beta}, \label{t11a}\\
  C_V  =  -k\beta^2\frac{\partial U}{\partial \beta}. \label{t11b}
\end{gather}
Thus, all we need is to calculate the partition function (\ref{t2})
using the corresponding energy spectrum.

Here we use both closed forms (\ref{osca}), (\ref{oscb}) and~(\ref{d9}), as well as the approximate forms (\ref{t1}) and~(\ref{d10}) of the spectra to calculate the canonical partition
function~(\ref{t2}) and to obtain thermodynamical quantities for the polymer solid and ideal gas.

In this work we have adopted the Maxwell--Boltzmann statistics for
particles with no spin. However, it has been argued that modif\/ied
statistics may be needed in quantum gravity and discrete theories.
The relation between the discrete structure and statistics is still
open~\cite{sw,chiapas}.

\subsection{Ensemble of \textit{polymer} oscillators}\label{section3.1}

\subsubsection{Exact spectrum}\label{section3.1.1}

To calculate the partition function using the spectrum~(\ref{osca}),
(\ref{oscb}), we split the sum into two parts
\begin{gather}\label{Zosc}
    Z(\beta) = \sum_n\!\exp\!\left[-\frac{\beta \hbar \omega}{\lambda^2}\left( 1 + \frac{\lambda^4}{8}a_n\left(\frac{4}{\lambda^4}\right) \right)
    \right] + \sum_{n'}\!\exp\!\left[-\frac{\beta \hbar \omega}{\lambda^2}\left( 1 + \frac{\lambda^4}{8}b_{n'+1}\left(\frac{4}{\lambda^4}\right) \right)
    \right]\!,\!\!\!\!\!
\end{gather}
the f\/irst term takes into account the even, and the second term the
odd, parts of the spectrum with $n$ and $n'$ running over all
integers.

To determine the thermodynamical quantities we f\/irst write the Helmoltz free energy as
\begin{gather*}
    F = -\frac{\mathcal{N}}{\beta} \ln \left\{ e^{-\frac{\beta \hbar \omega}{\lambda^2}} \sum_n \left[ \exp{\left(-\beta \hbar \omega
\frac{\lambda^2}{8}a_n\right)} + \exp{\left(-\beta \hbar \omega
\frac{\lambda^2}{8}b_{n+1}\right)}\right] \right\},
\end{gather*}
where we omit the argument $\frac{4}{\lambda^4}$ of the
characteristic Mathieu functions to avoid cumbersome expressions. From~(\ref{t9b}) and~(\ref{t9a})
it can be noticed that the equation of state and the chemical
potential remain unchanged with respect to the standard case. On one
hand, the Helmholtz free energy does not depend on the length of the
system, and, on the other, its dependence on $\mathcal{N}$ is not
modif\/ied with respect to the standard case.

\begin{figure}[t]
\centering
\subfigure[]{\includegraphics[width=2.8in]{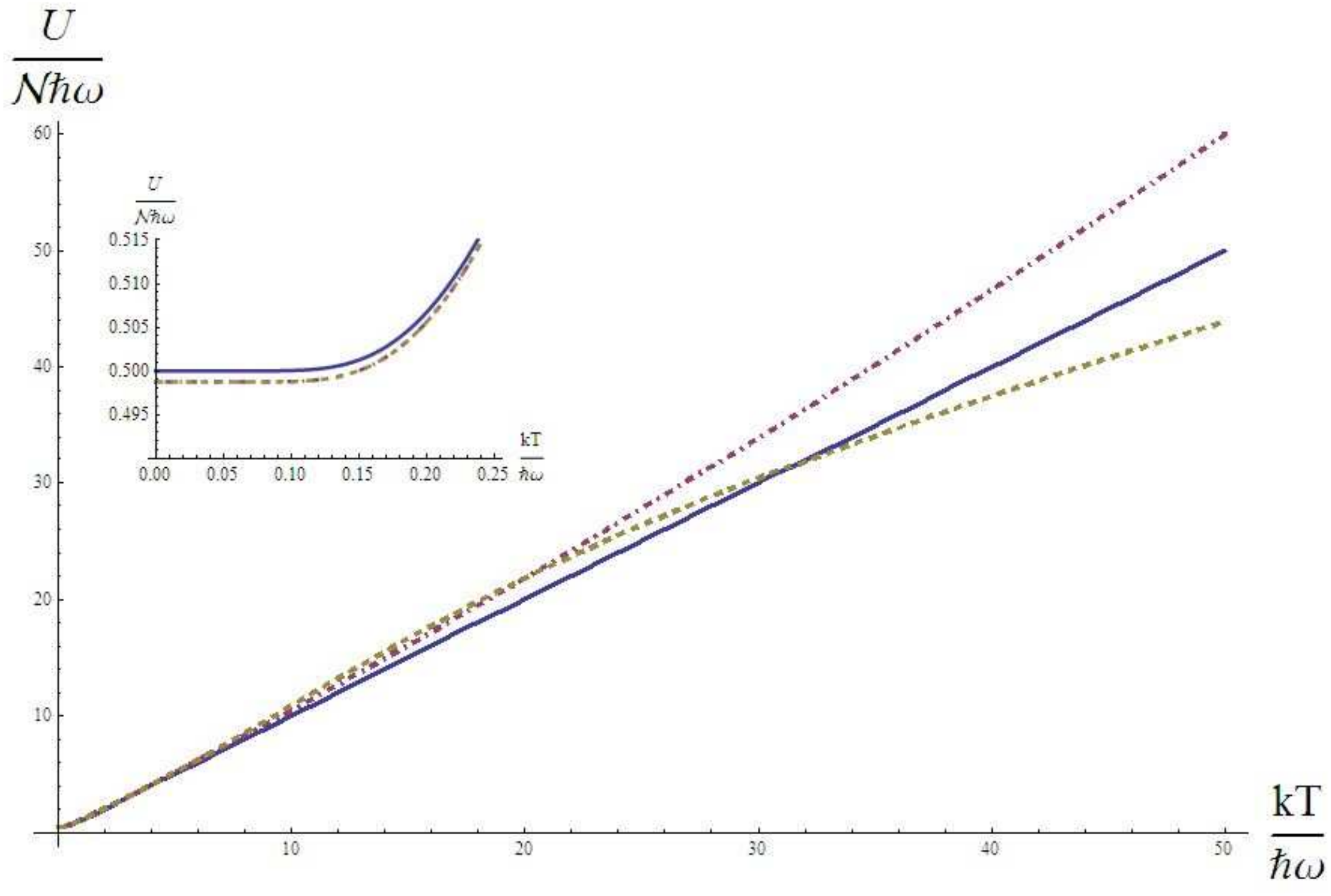}}\hfill
\subfigure[]{\includegraphics[width=2.7in]{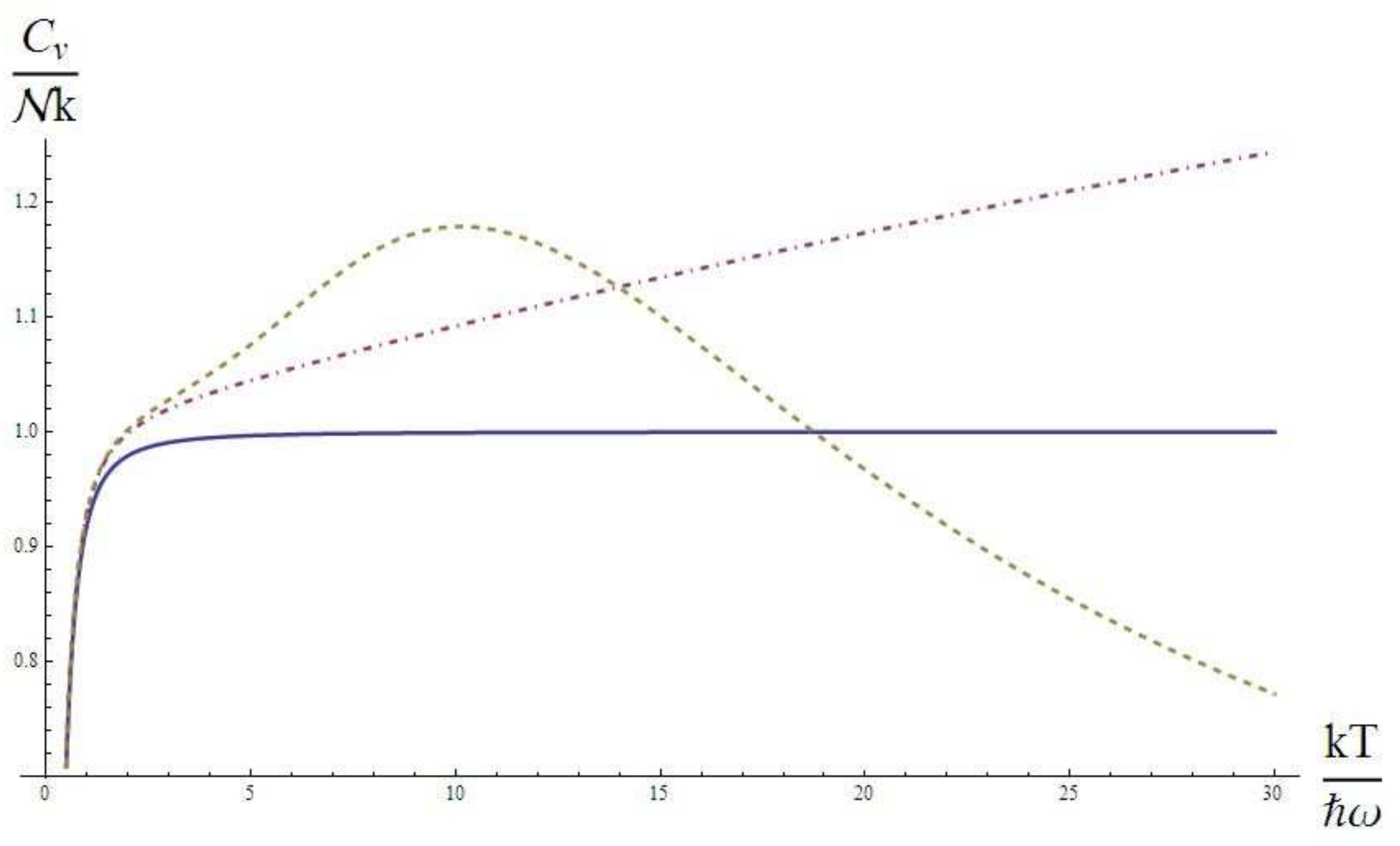}}

\caption{Here we plot the internal energy $U$ and the heat
capacity as functions of $kT/\hbar \omega$. In both graphics the
solid (blue) line corresponds to the standard case with $\lambda=0$,
the dashed (green) lines corresponds to
(\ref{Uoscpoly}) and (\ref{CVoscpoly}), respectively, for
$\lambda=0.2$. The dotdashed (red) lines correspond to the
approximated quantities (\ref{t12}) and (\ref{t13}) also for
$\lambda=0.2$. For the  internal energy $U$ we can see that the
ground energy is shifted in the inbox of (a) to a lower value, for
high temperatures the exact modif\/ied behavior decreases, while the
approximated one increases with respect to the standard case. For
the heat capacity $C_V$ we notice that for low temperature the three
cases are very similar (b), while for high temperatures the behavior
is completely dif\/ferent.} \label{Figs}
\end{figure}

The expressions for the entropy, internal energy and heat capacity
are the following:
\begin{gather}
  \frac{S}{\mathcal{N}k}  =  \ln{\left\{ e^{-\frac{\beta \hbar \omega}{\lambda^2}} \sum_n \Big[ e^{-\beta \hbar \omega
\frac{\lambda^2}{8}a_n} + e^{-\beta \hbar \omega
\frac{\lambda^2}{8}b_{n+1}}\Big] \right\}} \nonumber  \\
\phantom{\frac{S}{\mathcal{N}k}  = }{}  +    \frac{\beta \hbar \omega}{Z}\frac{e^{-\frac{\beta \hbar \omega}{\lambda^2}}}{\lambda^2} \sum_n\Big[A_n e^{-\beta \hbar \omega\frac{\lambda^2}{8}a_n} + B_n e^{-\beta \hbar \omega\frac{\lambda^2}{8}b_{n+1}} \Big], \nonumber 
\\
  U  =  \frac{\mathcal{N} \hbar \omega}{Z}\frac{e^{-\frac{\beta \hbar \omega}{\lambda^2}}}{\lambda^2} \sum_n\Big[A_n e^{-\beta \hbar \omega\frac{\lambda^2}{8}a_n} + B_n e^{-\beta \hbar \omega\frac{\lambda^2}{8}b_{n+1}} \Big], \label{Uoscpoly}\\
  \frac{C_V}{\mathcal{N}k}  =  \frac{e^{-\beta \hbar \omega\frac{1}{\lambda^2}}}{Z}\frac{(\beta \hbar \omega)^2}{\lambda^4} \Biggl\{    \sum_n\Biggl[A_n e^{-\beta \hbar \omega\frac{\lambda^2}{8}a_n}\left(A_n-\frac{2\lambda^2}{\beta \hbar \omega}\right)      \nonumber \\
  \hphantom{\frac{C_V}{\mathcal{N}k}  =}{}
  +   B_n  e^{-\beta \hbar \omega\frac{\lambda^2}{8}b_{n+1}} \left(B_n-\frac{2\lambda^2}{\beta \hbar \omega}\right) \Biggr]\! -\frac{e^{-\frac{\beta \hbar \omega}{\lambda^2}}}{Z}\left\{ \sum_n\Big[A_n e^{-\beta \hbar \omega\frac{\lambda^2}{8}a_n} + B_n e^{-\beta \hbar \omega\frac{\lambda^2}{8}b_{n+1}}
  \Big]\right\}^2 \!\nonumber \\
\hphantom{\frac{C_V}{\mathcal{N}k}  =}{}
    +  \frac{2\lambda^2}{\beta \hbar \omega} \sum_n\Big[
A_n e^{-\beta \hbar \omega\frac{\lambda^2}{8}a_n} + B_n e^{-\beta
\hbar \omega\frac{\lambda^2}{8}b_{n+1}} \Big] \Biggr\},
\label{CVoscpoly}
\end{gather}
where
\begin{gather*}
  A_n  \equiv  \left(\frac{\lambda^4}{8}a_n + 1\right), \qquad
  B_n  \equiv  \left(\frac{\lambda^4}{8}b_{n+1} + 1\right).
\end{gather*}
These sums can not be reduced to a simple form but they can be
treated numerically. These thermodynamical functions for arbitrary
$\lambda$ dif\/fer signif\/icatively with respect to the standard case
(see Fig.~\ref{Figs}). However, for small $\lambda$ the polymer and
standard cases behave qualitatively in a~similar manner.

Next we study the afore mentioned limiting cases. The case $\lambda
\ll 1$, corresponds to small deviations from the standard case
$\lambda =0$. Although the case $\lambda \gg 1$ has no direct
interpretation in terms of length scales, we can say that it
corresponds to the case when the energy of the system is much
greater than the energy associated with the polymer scale.

\subsubsection[Approximate spectrum $\lambda \ll 1$: ensemble of harmonic oscillators]{Approximate spectrum $\boldsymbol{\lambda \ll 1}$: ensemble of harmonic oscillators}\label{section3.1.2}

Using the approximation given in (\ref{t1}) for small $\lambda$,
consistent with $n_{\max}\sim\lambda^{-2}$, the partition function
becomes
\begin{gather*}
    Z(\beta)= \sum^{\lambda^{-2}}_{n=0} e^{-\beta \hbar \omega \left(\frac{1}{2}+n\right)}\exp{\left[\frac{\lambda^2}{16}\beta \hbar \omega  \left(\frac{1}{2}+n(n+1)\right)
    \right]},
\end{gather*}
summations can be performed in a closed form to yield
\begin{gather}\label{t6}
    Z(\beta)\simeq \frac{e^{-\frac{\beta \hbar \omega}{2}}}{1-e^{-\beta \hbar \omega}}\left[1 + \frac{\lambda^2}{32}\beta \hbar \omega \left(\frac{1+e^{-\beta \hbar \omega}}{1-e^{-\beta \hbar \omega}}\right)^2 +
    \mathcal{O}\big(e^{-\frac{1}{\lambda^2}}\big)\right].
\end{gather}
We recognize the f\/irst term in (\ref{t6}) as the usual partition
function of the standard oscillator; the second term is the leading
polymer correction and we are neglecting terms of order
$\mathcal{O}(e^{-\frac{1}{\lambda^2}})$ which clearly tend to zero
in the limit $\lambda\rightarrow 0$.

The regime where the f\/irst term is much larger than corrections,
i.e.\ the classical regime, corresponds to $\beta \hbar \omega \gg
1$. This can be seen in Fig.~\ref{Figs}. The polymer oscillator
approximates the standard oscillator for low temperatures.

Using the partition function (\ref{t6}) we calculate the
corresponding thermodynamic quantities, from~(\ref{t7}). Then we can
write a modif\/ied Helmholtz free energy
\begin{gather}
F = \frac{\mathcal{N}}{\beta}\left[\frac{\beta \hbar \omega}{2} + \ln{(1-e^{-\beta \hbar \omega})} - \ln{ \left(1 + \lambda^2\frac{\beta\hbar \omega}{32} \left(\frac{1+e^{-\beta\hbar
\omega}}{1-e^{-\beta\hbar \omega}}\right)^2 \right)} \right] \nonumber\\
\phantom{F}{}
   \cong \frac{\mathcal{N}}{\beta}\left[\frac{\beta \hbar \omega}{2} +
\ln{(1-e^{-\beta \hbar \omega})} - \lambda^2\frac{\beta\hbar
\omega}{32} \left(\frac{1+e^{-\beta\hbar \omega}}{1-e^{-\beta\hbar
\omega}}\right)^2
   \right].\label{t8}
\end{gather}
Substituting (\ref{t8}) into (\ref{t9a}), (\ref{t9c}) we obtain: $\mu
= F/\mathcal{N}$ and $p=0$: i.e.\ the equation of state does not change. As for the entropy we have
\begin{gather}\label{t10}
    S = \mathcal{N}k\left[\frac{\beta \hbar \omega}{e^{\beta \hbar \omega}-1} - \ln{(1-e^{-\beta \hbar \omega})} + \frac{(\lambda \beta \hbar \omega)^2}{8}e^{\beta \hbar \omega}\frac{e^{\beta \hbar \omega}+1}{(e^{\beta \hbar \omega}-1)^3}\right].
\end{gather}
For the internal energy we found
\begin{gather}\label{t12}
     U = \mathcal{N} \hbar \omega\left[\frac{1}{2} + \frac{1}{e^{\beta \hbar \omega}-1} -  \frac{\lambda^2}{32}
      \frac{(1+e^{\beta \hbar \omega})(e^{2\beta \hbar \omega} - 4e^{\beta \hbar \omega}\beta \hbar \omega-1)}{(e^{\beta \hbar \omega}-1)^3}  \right],
\end{gather}
and for the heat capacity
\begin{gather}\label{t13}
     C_V = \mathcal{N}k(\beta \hbar \omega)^2 \frac{e^{\beta \hbar \omega}}{(e^{\beta \hbar \omega}-1)^2}\left[1 +\frac{\lambda^2}{8}\left(\frac{2+\beta \hbar \omega(1+4e^{\beta \hbar \omega})+e^{2\beta \hbar \omega}(\beta \hbar \omega-2)}{(e^{\beta \hbar \omega}-1)^2}\right)\right].
\end{gather}
We can see from (\ref{t13}) that heat capacity is increased due to
the polymer correction, while it decreases the energy (\ref{t12}).
Since $\beta \hbar \omega \gg 1$, such modif\/ications are very small.

Physically, an ensemble of harmonic oscillators can be used to model
vibrations in solids (phonons). In a solid the vibrational modes are
modeled as a collection of harmonic oscillators in the so-called
harmonic approximation which consists of approximating the classical
interaction Hamiltonian of the atoms in a solid by a second order
Taylor series~\cite{path}. The simplest model is called the Einstein
model which assumes that all vibrational modes have the same
frequency $\omega$, and that the oscillators are independent with no
interaction. The thermodynamic magnitudes~(\ref{t8}), (\ref{t10}),
(\ref{t12}) and (\ref{t13}) contain modif\/ications to the
thermodynamics of an Einstein solid by def\/ining the vibrational
temperature $\Theta_V = \hbar \omega/k$. Note that in this case the
polymer scale is not involved in the def\/inition of vibrational
temperature.

\subsubsection[Approximate spectrum $\lambda \gg 1$: ensemble of rotors]{Approximate spectrum $\boldsymbol{\lambda \gg 1}$: ensemble of rotors}\label{section3.1.3}

\begin{figure}[t]
\centering
\includegraphics[width=3.5in]{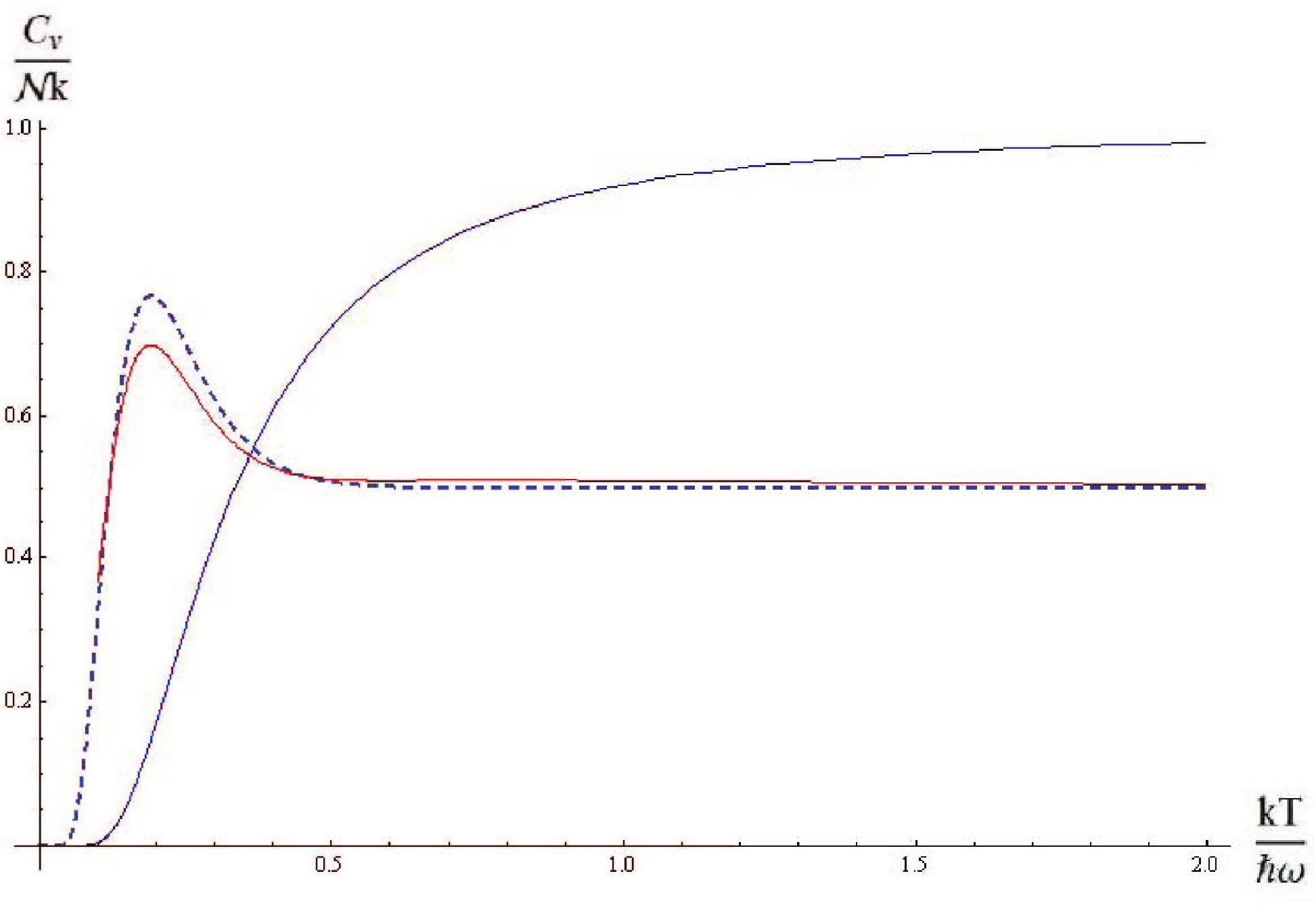}

  \caption{$C_V$ for the polymer oscillator. Blue line corresponds to the standard case with
$\lambda=0$, dashed line is the approximation using the partition
function (\ref{Zrot}) for $\lambda\gg 1$, and red line corresponds
to the case using the full spectrum in terms of Mathieu
characteristic functions in the same regime. We observe that, except
for the value of the maximum of~$C_V$, the approximation is very
good for most temperatures.} \label{FigCVLgde}
\end{figure}

Finally we consider the $\lambda \gg 1$ case for which the partition
function becomes
\begin{gather}\label{Zrot}
    Z(\beta) \cong  e^{-\frac{\beta \hbar \omega}{\lambda^2}} + e^{-\frac{\beta \hbar \omega}{\lambda^2}} \sum^{\infty}_{n=1}2  e^{-\frac{\beta \hbar \omega}{8}\lambda^2
    n^2} =  e^{-\frac{\beta \hbar \omega}{\lambda^2}} \vartheta_3\big(0, e^{-\frac{\beta \hbar
\omega}{8} \lambda ^2 } \big),
\end{gather}
where $\vartheta_3$ is the Jacobi's elliptic theta function~\cite{abrwtz}. As in previous cases, chemical potential and pressure
are unaf\/fected, while entropy internal energy and heat capacity are
as follows:
\begin{gather*}
  F  =  -\frac{\mathcal{N}}{\beta} \ln \big( e^{-\frac{\beta \hbar \omega}{\lambda^2}} \vartheta_3 \big),\\ 
  \frac{S}{\mathcal{N}k}  =  \ln \big(e^{-\frac{\beta \hbar \omega}{\lambda^2}}\,\vartheta_3\big)  +  \frac{\lambda^2 \beta \hbar \omega}{\vartheta _3} \left[ \frac{e^{-\frac{\lambda ^2 \beta \hbar \omega}{8}}\vartheta_3'}{8}  +  \frac{\vartheta _3}{\lambda^4}  \right], \\ 
  U  =   \mathcal{N}\hbar \omega \frac{\lambda^2}{\vartheta _3} \left[ \frac{e^{-\frac{\lambda ^2 \beta \hbar \omega}{8 }}\vartheta_3'}{8}     +     \frac{\vartheta _3}{\lambda^4}  \right], \\ 
  \frac{C_V}{\mathcal{N}k}  =  \frac{ (\beta \hbar \omega)^2}{64 \lambda^2} \left[   \frac{e^{-\frac{\lambda^2 \beta \hbar \omega}{4}}}{\vartheta _3^2}   \left(\lambda^6 \big( \vartheta _3\vartheta _3'' - \vartheta _3^{'2}\big)   +    e^{\frac{\lambda^2  \beta \hbar \omega}{8}} \left(\lambda^2-\frac{16}{ \beta \hbar \omega} \right) \vartheta _3\vartheta _3'\lambda^4    \right)  -  \frac{128}{ \beta \hbar \omega} \right] \nonumber \\
 \phantom{\frac{C_V}{\mathcal{N}k}=}{}
    + \frac{2  \beta \hbar \omega  \lambda^2}{\vartheta _3} \left[  \frac{e^{-\frac{\lambda ^2 \beta \hbar \omega}{8 }}\vartheta_3'}{8}  + \frac{\vartheta _3}{\lambda^4}  \right]. 
\end{gather*}
Again \looseness=-1
we omitted the arguments $0$ and $e^{-\frac{\beta \hbar
\omega}{8} \lambda ^2 }$ of $\vartheta_3$, while the primes in
$\vartheta_3'$ and $\vartheta_3''$, correspond to the f\/irst and
second derivatives of $\vartheta_3$ with respect to its second
argument $e^{-\frac{\beta \hbar \omega}{8} \lambda^2}$.
Notice from Fig.~\ref{FigCVLgde}, that the heat capacity of this
system tends asymptotically to its classical value at high
temperatures, while for low temperatures tends to zero, before
passing through a maximum. This same behavior is found in the
rotational contribution of particles with internal structure or an
ensemble of rotors~\cite{path}. Hence for
large $\lambda$ the quantum polymer oscillator approximates the
rotor.

Actually, one can def\/ine the corresponding rotational temperature
for this system. In the standard case the rotational temperature is
$\Theta_r \equiv \frac{\hbar^2}{2Ik}$, with $I$ the moment of
inertia.
Using in this case the ef\/fective moment of inertia
$I_{\rm ef\/f}=\frac{4\hbar}{\omega \lambda^2}$, the rotational
temperature turns out to be $\Theta_r \equiv \frac{\hbar \omega
\lambda^2}{8k} =\frac{m\omega^2\mu_0^2}{8k}$, which does depend on
$\lambda$. Note that the vibrational and rotational temperatures for
the system dif\/fer by a factor $\Theta_r/\Theta_V = \lambda^2/8$,
which indicates that for small~$\lambda$ the rotational states are
negligible, while the oscillatory states are for large $\lambda$.

\looseness=-1
The full behavior of the polymer system can be seen in Fig.~\ref{Figs}. The ensemble of polymer oscillators behave just as an
ensemble of quantum pendulums. For low temperature it approaches the
usual ensemble of oscillators, whereas at high temperatures it
approaches an ensemble of rotors. This behavior is evident when
studying the limiting cases separately as we have done here.

As usual in ensembles of systems from a spectrum containing $n^2$,
there is a maximum in~$C_V$ separating dif\/ferent behaviors. From
energy considerations, this peak appears due to a change of
concavity in~$U$, which can be seen in Fig.~\ref{Figs}(a). In the polymer oscillator as well as for the
pendulum, the maximum in~$C_V$ separates the rotational from the
oscillatory behavior as a~kind of smooth phase transition~\cite{naudts}. Given a f\/ixed $\lambda$ and for low temperatures,
the oscillatory states are turned on. As temperature increases
more energetic states appear and for the value of temperature at which~$C_V$ is maximum rotational states arise. For high temperatures, only the
rotational states remain excited. Moreover, it should be noticed
that the maximum of~$C_V$ depends on~$\lambda$. For $\lambda$ small
enough, the temperature at which the maximum occur, would be very
high so that the system approximates very well the oscillatory
behavior, while to reach the rotational behavior one would need a
huge amount of energy.

\subsection{Partition function for the \textit{polymer} ideal gas}\label{section3.2}

\subsubsection{Exact spectrum}\label{section3.2.1}

In the previous section we found that the energy spectrum of a
polymer quantum particle in a box is proportional to
$\cos{n\pi\lambda}$, namely equation~(\ref{d9}). We also showed that if we
expand $\cos{n\pi\lambda}$, when $\lambda \ll 1$, i.e., if the
dimensions of the box are large compared to the minimum scale~$\mu_0$, we obtain the usual quantum spectrum with modif\/ications of
order $\mathcal{O}(\lambda^2)$ equation~(\ref{d9}).

We can calculate the partition function with either the exact (\ref{d9}) or approximated spectrum~(\ref{d10}). Let us use the full spectrum (\ref{d9}). One way of
introducing the approximation $n\lesssim \lambda^{-1}$, for which
the analysis is valid, is to consider it as a cut-of\/f in the energy,
similarly as what has been done in~\cite{CVZ}. First we rewrite the
partition function as follows
\begin{gather}\label{t2.2}
    Z(\beta)= e^{- \frac{\Lambda^2}{2\pi\mu_0^2}} \sum^{\infty}_{n=0}\exp{\left(\frac{\Lambda^2}{2\pi\mu_0^2}\cos{n\pi\lambda}
    \right)},
\end{gather}
where $\Lambda = \sqrt{\frac{2\pi\beta \hbar^2}{m}}$ is the thermal
wave length \cite{path}, and the exponential in the argument of the
sum can be written conveniently as an inf\/inite sum of modif\/ied
Bessel functions of f\/irst order~\cite{abrwtz}
\begin{gather}\label{bess0}
    e^{\frac{\Lambda^2}{2\pi\mu_0^2}\cos{n\pi\lambda}} = I_0\left(\frac{\Lambda^2}{2\pi\mu_0^2}\right) + 2\sum_{k=1}^{\infty} I_k\left(\frac{\Lambda^2}{2\pi\mu_0^2}\right) \cos{(k n \pi\lambda)}.
\end{gather}
Then we can replace (\ref{bess0}) in (\ref{t2.2}) and perform the
sum over $n$. However, we immediately see that the f\/irst term
diverges. It is then necessary to consider the sum only up to
$1/\lambda$, as dictated by the approximation. Moreover, when
$\lambda\rightarrow0$, the limit of the sum tends to $\infty$, as in
the standard case. Thus, the sum in~(\ref{t2.2}) becomes
\begin{gather*}
    \sum^{1/\lambda}_{n=0}\exp \!\left(\frac{\Lambda^2}{2\pi\mu_0^2}\cos{n\pi\lambda}\right)\!
    = \frac{1}{\lambda}I_0\!\left(\frac{\Lambda^2}{2\pi\mu_0^2}\right)\! + \cosh\!\left(\frac{\Lambda^2}{2\pi\mu_0^2}\right) \! + \sum_{k=1}^{\infty} I_k\!\left(\frac{\Lambda^2}{2\pi\mu_0^2}\right)\! \cot \! \left(\frac{k\pi \lambda}{2}\right) \sin{k \pi}.\!
\end{gather*}
In the last term $\cot{k\pi \lambda/2} \sim 2/(k\pi\lambda)$ for
small $\lambda$, then it is zero for any integer $k$. The partition
function for this case has the form:
\begin{gather}\label{bess3-0}
    Z(\beta) = I_0\left(\frac{\Lambda^2}{2\pi\mu_0^2}\right)\frac{e^{- \frac{\Lambda^2}{2\pi\mu_0^2}} }{\lambda} + \frac{1}{2}\Big(1+e^{-\frac{\Lambda^2}{\pi\mu_0^2}}\Big).
\end{gather}
Let us notice that the last term $e^{-\frac{\Lambda^2}{\pi\mu_0^2}}$
tends to zero as $\mu_0\rightarrow 0$, also there is a constant term~$1/2$, which only redef\/ines the scale of $Z$\footnote{Moreover, one
can think that this term comes from the way we approximate the sum.
We can realize this by using the Euler--Maclaurin formula to
calculate the sum in the partition function, this calculation is
usual in standard statistical mechanics
\[\sum^{N}_{n=0}f(n) \cong \int^{N}_{0}f(n)dn + \frac{1}{2}(f(0)+f(N))+\cdots,\]
using $N=1/\lambda$ and $f(n) =
\exp{\left(\frac{\Lambda^2}{2\pi\mu_0^2}\cos{n\pi\lambda}\right)}$
we f\/ind, as we shall see, that the partition function coincides with
the asymptotic expansion of (\ref{bess3-0}) and also contains the
factor $1/2$, then it is not due to polymer corrections and we can
ignore it from now on.}. Consider then the partition function as
\begin{gather*}
    Z(\beta) = I_0\left(\frac{\Lambda^2}{2\pi\mu_0^2}\right)\frac{e^{- \frac{\Lambda^2}{2\pi\mu_0^2}} }{\lambda},
\end{gather*}
and let us calculate the corresponding thermodynamical quantities by
recalling that the Helm\-holtz free energy in the case of
indistinguishable particles can be determined by the relation
$F=-\beta^{-1}\ln{\left(\frac{Z^{\mathcal{N}}}{\mathcal{N}!}\right)}$,
\cite{path}, which in terms of $\beta$ reads
\begin{gather*}
   F = -\frac{\mathcal{N}}{\beta}\left[ 1-\ln{\mathcal{N}}+\ln{\left(\frac{L}{\mu_0}I_0\, e^{- \frac{\Lambda^2}{2\pi\mu_0^2}}\right)} \right],
\end{gather*}
the other thermodynamical quantities are as follows:
\begin{gather*}
   \mu = \frac{1}{\beta}\left[ \ln{\mathcal{N}} - \ln{\left(\frac{L}{\mu_0}I_0\, e^{- \frac{\Lambda^2}{2\pi\mu_0^2}}  \right)} \right],
\qquad
     p = \frac{\mathcal{N}}{\beta L},
\\
    S = \mathcal{N}k \left[ 1 -\ln{\mathcal{N}} + \ln{\left(\frac{L}{\mu_0}I_0\, e^{- \frac{\Lambda^2}{2\pi\mu_0^2}} \right)} + \frac{\hbar^2 \beta}{m \mu^2_0}\left(1 - \frac{I_1}{I_0} \right)\right],
\\
     U = \frac{\mathcal{N} \hbar^2}{m\mu^2_0} \left(1-\frac{I_1}{I_0} \right),
\qquad
     C_V = \frac{\mathcal{N}k}{2}\left(\frac{\hbar^2 \beta}{m\mu^2_0} \right)^2  \frac{I_0\left( I_0 + I_2 \right) -
     2I_1^2}{I_0^2}.
\end{gather*}
In all previous expressions we omit the arguments of the Bessel
functions. We notice that also in this case, with the assumptions
that we made, the equation of state remains unchanged.

\begin{figure}[t]
\centering
 \includegraphics[width=3.8in]{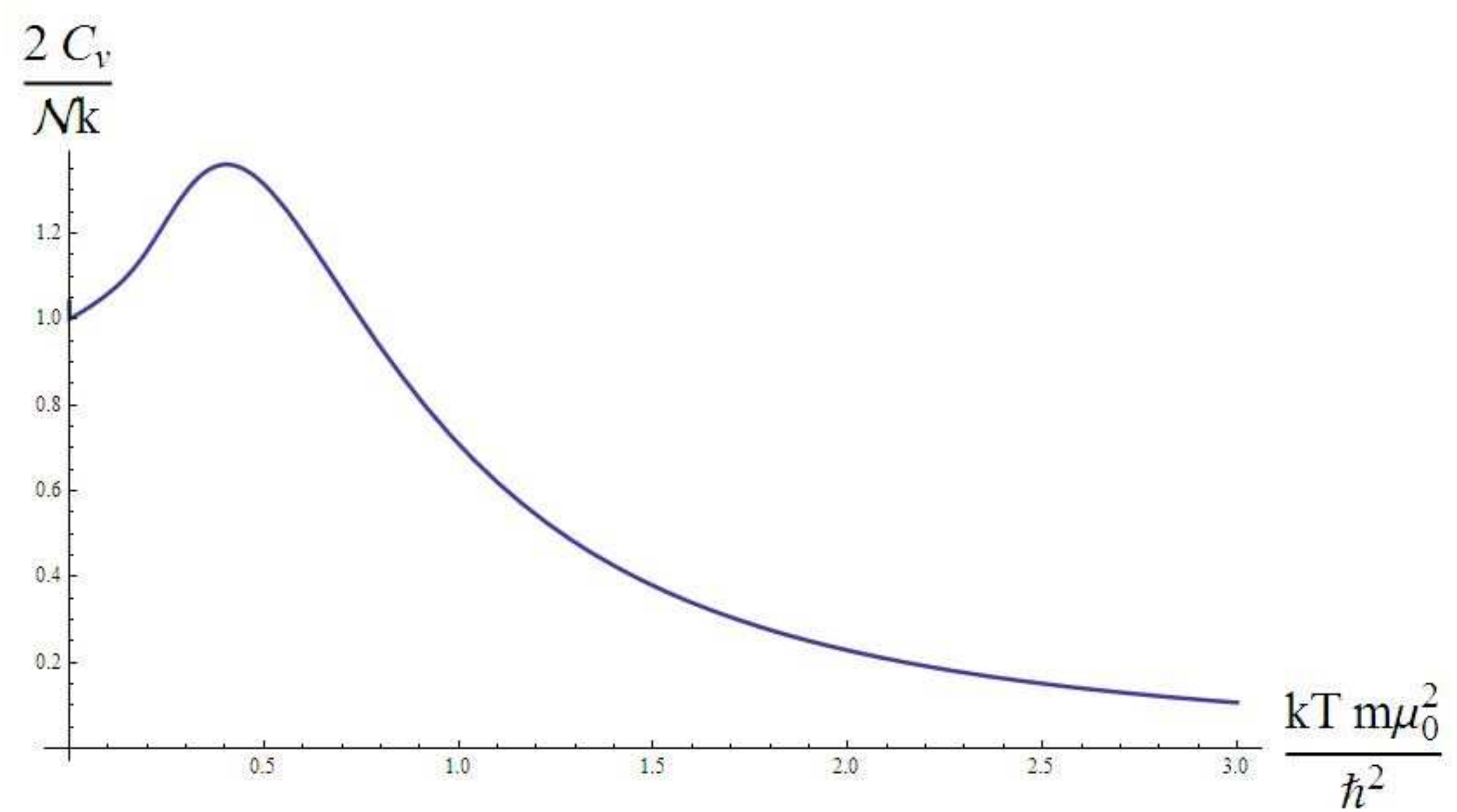}

  \caption{The $C_V$ for the polymer ideal gas has a maximum around $T \sim 2\Theta_{\rm poly}$,
  for smaller values of temperature one recovers the usual behavior, and when $T\rightarrow\infty$, $C_V$ tends to zero.}
  \label{FigCVGasPoly}
\end{figure}

From Fig.~\ref{FigCVGasPoly}, we notice that for low temperature
(with generic $\mu_0$) $C_V$ features the usual ideal gas behavior
namely, it takes a constant value. On the other hand as temperature
increases $C_V$ reaches a maximum (which depends on $\mu_0$) and
then goes to zero as $T\rightarrow\infty$. This is consistent with
the asymptotic behavior of the energy in the same regime.
The maximum of $C_V$ occurs at about $T \sim \Theta_{\rm poly}/2$, where
$\Theta_{\rm poly} \equiv \frac{\hbar^2}{km\mu_0^2}= \frac{E_*}{k}$.
Remarkably, since the energy of the polymer particle in the box is
bounded between 0 and $E_*$, we are just regaining the so called
\emph{Schottky effect},~\cite{path}, that appears for a two level
quantum system in which the peak of~$C_V$ appears for a~temperature
given by the energy dif\/ference between levels divided by~$k$.

Moreover, it is well known that systems which are bounded from
above, as the present case, allow the existence of negative
temperatures \cite{Rams}. From the thermodynamical def\/inition of
temperature, negative values of $T$ correspond to negative values in
the slope of the graph of energy versus entropy. To see how this may
happen one notices that in the partition function~(\ref{t2}), and
for positive temperatures, higher energy states contribute less than
the low energy ones. this situation gets reversed for negative
temperatures and in this situation it is mandatory that the energy
be bounded above for the partition function to make sense. That's
why only systems such as those having two levels as magnetic systems
and nuclear spin systems, feature negative temperatures~\cite{Rams}.
Negative temperatures correspond to energies that are in principle
experimentally accessible, due to the fact that $C_V$ remains
positive for negative temperatures, e.g.\ in nuclear spin systems~\cite{Rams}.

\begin{figure}[t]
\centering
\includegraphics[width=3.8in]{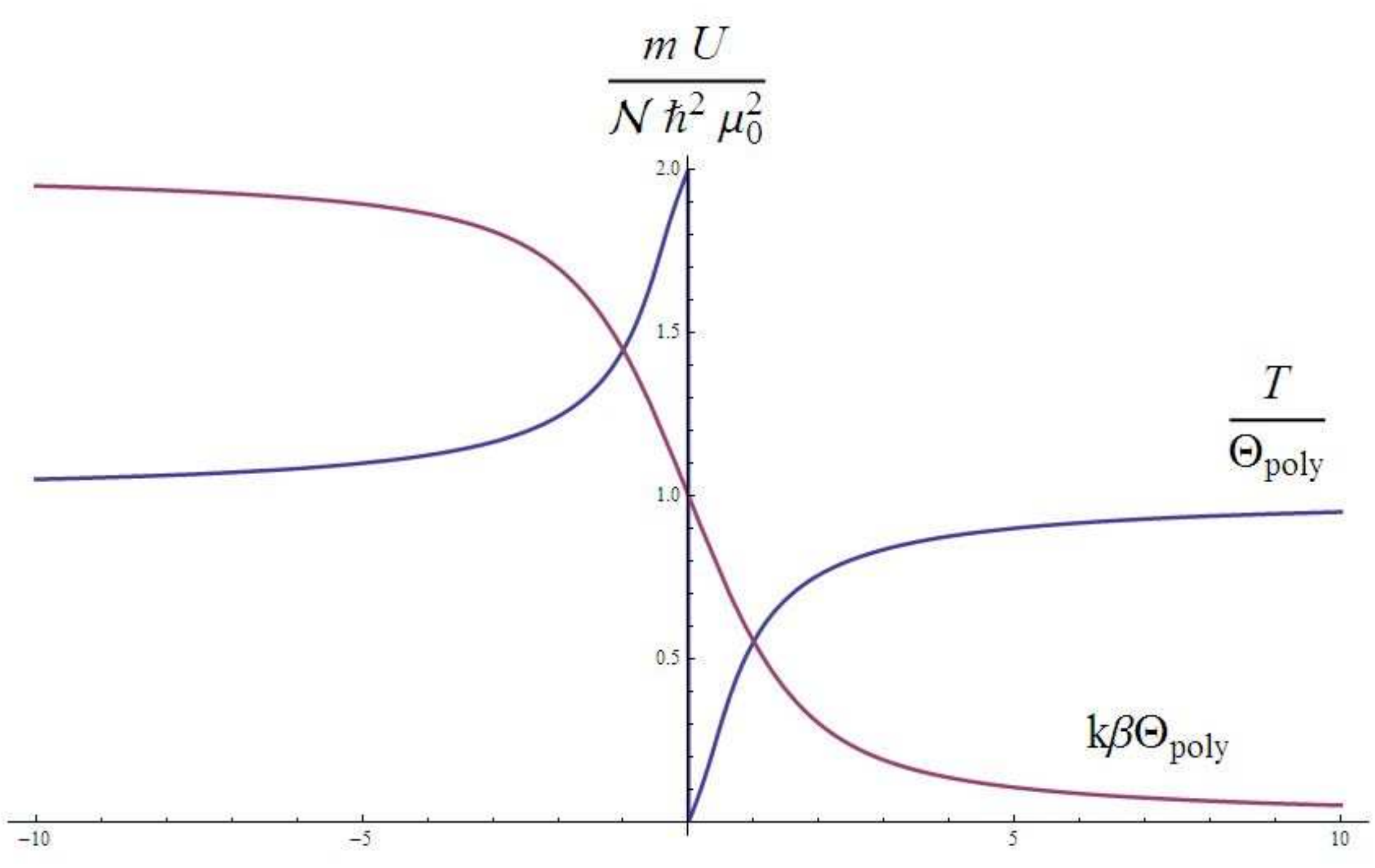}

  \caption{Internal energy of the polymer ideal gas as function of temperature $T$ and its reciprocal $\beta$
  considering negative values of them.}  \label{FigUGasPoly}
\end{figure}

Now let us analyze our polymeric gas in the region of negative
temperatures. In the limit  $T\rightarrow 0^+$, $U$ reaches its
minimum value which is zero; however, in the limit $T\rightarrow
0^-$, $U$ takes a~value that is twice its asymptotic value. In the
positive temperature region the energy increases from zero to its
maximum value $\frac{\mathcal{N}\hbar^2}{m\mu_0^2}$, as
$T\rightarrow\infty$. In the negative temperatures regime~$U$
decreases from twice to once its asymptotic value as
$T\rightarrow-\infty$. We can see this behavior from the Fig.~\ref{FigUGasPoly}.
In the polymer case, when $\mu_0$ is very small, the maximum energy
for positive temperature tends to inf\/inity and we can not reach the
negative temperature regime. To access negative temperature in the
polymer case will be very dif\/f\/icult in the case~$\lambda\ll 1$.

\subsubsection{Approximate spectrum}\label{section3.2.2}

Let us consider the limit $\lambda \rightarrow 0$ of the partition
function (\ref{bess3-0}) in order to regain the results for the
standard ideal gas. We can make use of the asymptotic expansion of
$I_0$ \cite{abrwtz}, which yields:
\begin{gather}\label{bess4}
    Z(\beta) \approx \frac{L}{\Lambda} \left[1 + \frac{\pi}{4} \frac{\mu_0^2}{\Lambda^2}+ \frac{9\pi^2}{32} \frac{\mu_0^4}{\Lambda^4}+ \frac{75\pi^3}{128} \frac{\mu_0^6}{\Lambda^6}+ \frac{3675 \pi^4}{2048} \frac{\mu_0^8}{\Lambda^8}  +  \cdots \right] + \frac{1}{2}\Big(1+e^{-\frac{\Lambda^2}{\pi\mu_0^2}}\Big).
\end{gather}
Of the two terms in round brackets in (\ref{bess4}) only the $1/2$
remains in the $\mu_0\rightarrow 0$ limit. However, it only yields a
constant shift in $Z$.

Interestingly, a dif\/ferent approximation from the above is to use
the approximated spectrum~(\ref{d10}) instead of~(\ref{d9}) in the
partition function. Since $\lambda \ll 1$  we consider only the
f\/irst terms in the series of the second exponential, so that the
partition function could be expressed~as
\begin{gather}\label{t15}
    Z(\beta) \cong \sum_{n=0}^{\infty}e^{-\frac{\beta \hbar^2\pi^2}{2mL^2}n^2} + \lambda^2\frac{\beta \hbar^2 \pi^4}{24mL^2}\sum_{n=0}^{\infty} n^4 e^{-\frac{\beta \hbar^2\pi^2}{2mL^2}n^2} + \mathcal{O}\big(\lambda^4\big).
\end{gather}
Now, by the use of the Poisson resummation formula, namely
\begin{gather*}
    \sum_{n=-\infty}^{\infty}f(n) = \sum_{n=-\infty}^{\infty}\int_{-\infty}^{\infty} f(y)e^{-2\pi i yn}
    dy,
\end{gather*}
we can write the partition function as
\begin{gather*}
    Z(\beta) \cong  \sqrt{\frac{mL^2}{2\pi\beta \hbar^2}} + \lambda^2 \frac{\pi}{4} \left(\frac{mL^2}{2\pi\beta \hbar^2} \right)^{3/2}  + \cdots +  \mathcal{O}\Big(e^{-\frac{2mL^2}{\beta \hbar^2
    }}\Big),
\end{gather*}
that is precisely the f\/irst term in (\ref{bess4}). The f\/irst term is
the standard partition function for the one dimensional ideal gas.
This leads us to the constraint for the approximation to hold, that
$\mu_0 \ll \Lambda\sqrt{\frac{4}{\pi}}$.

Furthermore we can compute it with more orders in $\mu_0$
\begin{gather}\label{t19.12}
    Z(\beta) \cong \frac{L}{\Lambda} \left[1 + \frac{\pi}{4} \frac{\mu_0^2}{\Lambda^2}+ \frac{9\pi^2}{32} \frac{\mu_0^4}{\Lambda^4}+ \frac{75\pi^3}{128} \frac{\mu_0^6}{\Lambda^6}+ \frac{3675 \pi^4}{2048} \frac{\mu_0^8}{\Lambda^8}  +  \mathcal{O}\Big(e^{-4\pi\frac{L^2}{\Lambda^2}},
    \mu_0^{10}\Big)\right].
\end{gather}
Let us use (\ref{t19.12}) to order $\mu_0^2$ to obtain
the approximate thermodynamic quantities:
\begin{gather*}
    F = \frac{-\mathcal{N}}{\beta}\left[1 - \ln{\mathcal{N}} + \ln{L} - \frac{1}{2}\ln{\beta} + \ln{\sqrt{\frac{m}{2\pi \hbar^2}}} +\ln{\left(1+\frac{\mu_0^2}{8}\frac{m}{\beta\hbar^2} \right)} \right].
\end{gather*}
We notice that $F$ diverges for $\beta=0$ or $T\rightarrow \infty$,
but with the opposite sign than in the usual case. As we approach to
$\beta=0$ we found a maximum after which the free energy diverges to~$-\infty$ at $\beta=0$. This means that the ef\/fect of the
minimum length scale on the free energy, is to provide a turning
point for $F$ at very high temperatures. Furthermore, when we
increase the value of~$\mu_0$, the free energy becomes negative at
high temperatures.

We can calculate gas pressure through the relationship $p = -
\partial F / \partial L $, which is valid for one dimensional
system. Realizing that the term that comes from the polymer
correction does not depend on $L$, but on~$ \mu_0 $, we can obtain
the equation of state for the ideal gas as $ p = N / \beta L $,
which is the same as in the continuum case. The polymer corrections
can not be seen as an ef\/fective interaction among the particles as
it was the case for GUP or MDR~\cite{camacho}.

Recalling (\ref{t9b}) and (\ref{t9c}) the chemical potential and the
entropy of the gas can be calculated
\begin{gather}
    \mu = -\frac{1}{\beta}\left[- \ln{\mathcal{N}} + \ln{L} - \frac{1}{2}\ln{\beta} + \ln{\sqrt{\frac{m}{2\pi \hbar^2}}} +\ln{\left(1+\frac{\mu_0^2}{8}\frac{m}{\beta\hbar^2} \right)} \right],
\nonumber\\
\label{t23}
    S = \mathcal{N}k \left[\frac{3}{2} +\ln{\frac{L}{\mathcal{N}}}-\ln{\beta^{1/2}}+\ln{\sqrt{\frac{m}{2 \pi \hbar^2 }}}+ \frac{\mu_0^2 m}{8\hbar^2 \beta} + \ln{\left(1 + \frac{\mu_0^2m}{8\hbar^2 \beta}\right)} \right].
\end{gather}
The last expression (\ref{t23}) would be the equivalent of a one
dimensional \emph{Sackur--Tetrode's} formula for the entropy with
corrections due to the underlying discreteness. The energy and heat
capacity are obtained from (\ref{t11a}) and (\ref{t11b})
respectively as
\begin{gather}
    U = \frac{\mathcal{N}}{2 \beta} \left(1 + \frac{\mu_0^2}{4} \frac{m}{\hbar^2 \beta} + \cdots \right),
\nonumber\\
\label{t25}
    C_V = \frac{k \mathcal{N}}{2}  \left(1 + \frac{\mu_0^2}{2}\frac{m}{\hbar^2 \beta} + \cdots
    \right).
\end{gather}
All the above expressions are also obtained when considering the
asymptotic expansion of Bessel functions of the quantities obtained
in the previous section.

We notice from (\ref{t25}) that heat capacity has a dif\/ferent
behavior with respect to the standard case for which $C_V $ is
constant. For $\beta=0$, or high temperature, indeed it diverges in
this approximate case. However, we know that this is only an
approximation which corresponds, in the full case, to increasing
$C_V$ to its maximum. Because heat capacity is related to energy
f\/luctuations, we can say that for high temperatures there are strong
f\/luctuations in the energy due to space discreteness.

Notice that if we use the result (\ref{d12}) the calculations of the
thermodynamical quantities remain the same and only dif\/fer by
numerical factors, i.e.\ in the partition function (\ref{t15}) a
factor~$1/24$ is replaced by $-1/3$, etc. The main dif\/ference is
that with (\ref{d12}) the thermodynamical quantities will decrease
instead of increase as our results show.

\section{Discussion}\label{section4}

Polymer quantum mechanics considers mechanical
models quantized in a similar way as loop quantum gravity but in
which loops/graphs resembling polymers are replaced by discrete sets
of points. It has allowed to study some features of loop quantum
gravity in a simpler context, namely through the use of mechanical
systems \cite{ashpqm,CVZ1}. Indeed this opened up the possibilities
to investigate dif\/ferent physical problems
\cite{ashpqm,chiou,PUR,Huss,Kunst}. On the other hand important
gravitational systems like the cosmos itself and black holes
necessarily involve thermodynamics in their description. However,
little attention has been given to the thermostatistics of polymer
quantum systems. In this work we embarked on this task using the
canonical ensemble theory applied to a polymer solid and a polymer
gas, both in one dimension. The resulting thermodynamic quantities
have modif\/ications due to the minimum length scale that introduces
the polymer quantization. Thermodynamics with modif\/ications due to
quantum gravity has been studied in the context of extensions of the
standard model with Lorentz violations \cite{don} and also models
that mimic the graphic states of loop representation, but which,
however, do not represent physical situations \cite{qgrphy}. It is
important to stress that here the canonical partition function is
used as the simplest case, the relation between the discrete space
and the statistics is an open issue \cite{sw,chiapas}.

First we consider an ensemble of polymer oscillators, and we noticed
that it can be interpreted as an ensemble of quantum pendulums with
two regimes: For small $\lambda$ one has the vibrational or simple
oscillator regime, and for large $\lambda$ the rotational regime.
Both behaviors are separated by a maximum in the $C_V$. In the
generic $\lambda$ case, we observe that to access the rotational
regime requires a high temperature that also depends on $\lambda$.
In that case the partition function and therefore the thermodynamic
quantities, are written in terms of certain sums of the exponentials
containing the Mathieu characteristics functions (\ref{Zosc}).

In the vibrational regime $\lambda \ll 1$ the partition function can
be expressed as a power series on the minimum polymer length scale.
Interestingly the equation of state is not modif\/ied. As is well
known, with this model one can model an Einstein's solid introducing
the vibrational temperature $\Theta_V$ that takes into account the
characteristic energy of the system in this regime. We recover the
known thermodynamical quantities when $\lambda \rightarrow 0$.

The rotational regime $\lambda \gg 1$ that corresponds to an
ensemble of rotors, can be characterized by the rotational
temperature $\Theta_r$ that in this case depends on $\lambda$, as
$\Theta_r = \frac{\hbar \omega}{8k}\lambda^2$. In this case $Z$ and
the thermodynamical variables can be written in terms of the Jacobi's
theta function~$\vartheta_3$,~(\ref{Zrot}). This regime can be of
interest for the polymer analogous to the trans-planckian problem,
since this case corresponds to energies beyond the characteristic
polymer energy given by the coef\/f\/icient of equation~(\ref{3.1}).

Amusingly it is possible to consider the ef\/fect of the polymer quantization in the case of electromagnetic radiation as follows.
Let us recall that an ensemble of oscillators
can be used to model excitations in solids (phonons), but also
quantum excitations of the electromagnetic f\/ield (photons).
Historically, equilibrium radiation was f\/irst studied by Planck in
1900, who consi\-de\-red this system as an ensemble of harmonic
oscillators with the same frequency. Nowadays we consider that
photons are ultra relativistic bosons with some particular energy
$\hbar \omega$~\cite{path}. Let us use Planck's simple model. By
interpreting the internal energy as $\mathcal{N}$ times the average
energy of each oscillator, i.e.\  $U=\mathcal{N}\langle E_n \rangle$,
from which it follows that $\langle E_n \rangle = \frac{\hbar
\omega}{2} + \langle n \rangle$, where $\langle n \rangle$ is the
average occupation number that can be obtained from the previous
equation and~(\ref{t12}). This yields
\begin{gather}\label{bb0}
    \langle n \rangle = \frac{1}{e^{\beta \hbar \omega}-1} - \frac{\lambda^2}{32}\frac{(1+e^{\beta \hbar \omega})(e^{2\beta \hbar \omega} - 4\beta \hbar \omega e^{\beta \hbar \omega} - 1)}{(e^{\beta\hbar
    \omega}-1)^3},
\end{gather}
where the second term is the polymer correction. With (\ref{bb0}) we
can calculate the spectral density that is def\/ined as
\begin{gather}\label{bb1}
    u_{\omega} \equiv \frac{d}{d\omega}\left(\frac{U}{V}\right) = \langle n
\rangle\,\hbar \omega  \frac{g(\omega)}{V},
\end{gather}
where the density of states $g(\omega) =
\frac{V\omega^2}{\pi^2c^3}$, $V$ being the volume. The expression
(\ref{bb1}) is thus the black body distribution now containing
polymer corrections. Note that the spectral density for a f\/ixed
temperature is modif\/ied for high frequencies. We know that the
cosmic background radiation CMB with $T\cong 3K$ is the black body
that has been measured more accurately \cite{smoot}. This further
constrains the possible value of $\lambda$ and therefore the minimum
scale of $\mu_0$. In this case such changes do not alter the
functional dependence on temperature as in other approaches
\cite{nozsef,camacho,ipny}. Using (\ref{bb0}) and
(\ref{bb1}) we can obtain directly the energy density by
integration. It follows that the Stefan--Bolztmann law takes the form
\begin{gather}\label{bb2}
     \frac{U}{V} = \frac{\pi^2 k^4}{15c^3\hbar^3} \left( 1 + \lambda^2 \frac{135}{4 \pi^4}\zeta(3)
    \right)T^4 \simeq \frac{\pi^2 k^4}{15c^3\hbar^3} \left( 1 +
    0.416485  \lambda^2 \right)T^4.
\end{gather}
Thus, given (\ref{bb2}) the Stefan--Boltzmann constant is modif\/ied by
a term of order $\mathcal{O}(\lambda^2)$ as
\begin{gather*}
    \sigma_{_{SB}}^{(\lambda)} = \frac{\pi^2 k^4}{60 c^2\hbar^3} \left( 1
+   0.416485 \lambda^2\right).
\end{gather*}
where $\lambda$ is small. Notice that, as expected, we recover the
usual formulae of thermodynamics of  black body radiation for
$\lambda \rightarrow 0$. Similar analysis using other proposal have
been given in~\cite{nozsef,camacho,ipny}.

As for the ideal polymer gas, the partition function  was determined in
two regimes depending on whether the spectrum is considered in its
exact or approximate form. In the exact spectrum case we notice
there is a cut-of\/f in the energy levels proportional to~$1/\lambda$.
This ensures in particular the convergence of the partition function~(\ref{t2.2}). We note that in this system is quite evident that
negative temperatures are allowed. In the approximate
spectrum case we express the partition function as a series in
powers of~$\mu_0$. Of course standard thermodynamics is contained in
our results in the limit $\lambda\rightarrow 0$. These results resemble those
obtained from GUP~\cite{nozmeh,nozazizi} in the sense they correspond to quadratic corrections although they feature opposite tendencies~\cite{taub}. Clearly within our analysis the origin of the
modif\/ications can be traced all the way back to the polymer model,
in which there is also a modif\/ied uncertainty relation \cite{PUR}.

In regard to the thermodynamics of both of our systems it is worth
stressing a common behavior of their heat capacity. For the gas, in
contrast to the standard case, the~$C_V$ has a~maximum of about half
of $\Theta_{\rm poly}$, the temperature at which the polymer ef\/fects are
evident. The polymer oscillator, on the other hand, also shows a~similar behavior in its $C_V$, having a~maximum that separates two
dif\/ferent behaviors characterized by~$\Theta_V$ and~$\Theta_r$.

Now we mention some possible extensions of this work. The canonical
distribution adopted here is only intended to be an
approximation. The statistics of polymer systems, which are
naturally discrete is still an open issue \cite{sw}. In
\cite{chiapas} the problem of calculating the number of accessible
microstates was considered in a semiclassical perspective; it was
found that the standard methods yield only approximate results. To
count states in polymer systems and to include fermions and bosons,
again a better understanding of the statistics of polymer quantum
systems is needed.

In this work we learnt that polymer quantization induces some ef\/fects
on the statistical description of the systems, which in principle we
can explore. This ef\/fects are  such as the maximum in $C_V$ between
two dif\/ferent behaviors, or the appearance of negative temperatures
in the case of ideal gas.
Interestingly, a similar behavior has been observed in gravitational
systems such as black holes, where the heat capacity shows a phase
transition that depends on the minimum length scale~\cite{Husain-Mann}. The study of the thermodynamic quantities of
gravitational systems has focused mainly on obtaining the entropy of
black holes from gravitational quantum states~\cite{BHE}, or to f\/ind
ef\/fective modif\/ications to those thermodynamical quantities,~\cite{Poly1,tlqc}. However, recently there has been some
quantum gravity models which are based on thermostatistics of
condensed matter systems, can yield interesting results~\cite{Eugenio,Fotini,Fotini2}.

Certainly an interesting question emerges when exploring the
thermodynamics of non equilibrium processes such as those that occur
in the early universe or in the black hole evaporation. Clearly it
is necessary to extend the methods presented here to address these
problems.

\subsection*{Acknowledgements}

We would like to thank A.~Camacho, M.~Reuter, J.A.~Zapata and E.~Flores for useful
discussions, comments and suggestions. We are indebted to Viqar Husain for communicating us recent results related to the present work~\cite{Husainetal}.
Partial support from the
following grants is acknowledged: CONACyT-NSF Strong
backreaction ef\/fects in quantum cosmology, CONACyT 131138 and
SNI-III research assistant 14585 2866 (GCA),  DAAD A/07/95322 and
CONACyT 55521 (EM).

\pdfbookmark[1]{References}{ref}
\LastPageEnding

\end{document}